# Meta-analysis models relaxing the random effects normality assumption: methodological systematic review and simulation study

Kanella Panagiotopoulou[1], Theodoros Evrenoglou[2], Christopher H Schmid[3], Silvia Metelli[1], Anna Chaimani[1,4]

[1]Université Paris Cité, Center of Research in Epidemiology and Statistics, Inserm, Paris, France

[2]Institute of Medical Biometry and Statistics, Faculty of Medicine and Medical Center-University of Freiburg, Freiburg im Breisgau, Germany

[3]Department of Biostatistics, School of Public Health, Brown University, Providence, RI, USA

[4]Oslo Center for Biostatistics and Epidemiology, Department of Biostatistics, University of Oslo, Oslo, Norway

Corresponding author:

Kanella Panagiotopoulou

Center of Research in Epidemiology and Statistics (CRESS), METHODS Team

Université Paris Cité, Inserm

Hôpital Hôtel-Dieu, 1 Place du Parvis Notre-Dame

Paris 75004, France

email: **kanella.panagiotopoulou@etu.u-paris.fr**

**Data availability**

R code for the simulation study is available from https://github.com/Kanella-web/Non-normal-meta-analysis-models.git

**Abstract**

Random effects meta-analysis is widely used for synthesizing studies under the assumption that underlying effects come from a normal distribution. However, under certain conditions the use of alternative distributions might be more appropriate. We conducted a systematic review to identify articles introducing alternative meta-analysis models assuming non-normal between-study distributions. We identified 27 eligible articles suggesting 24 alternative meta-analysis models based on long-tail and skewed distributions, on mixtures of distributions, and on Dirichlet process priors. Subsequently, we performed a simulation study to evaluate the performance of these models and to compare them with the standard normal model. We considered 22 scenarios varying the amount of between-study variance, the shape of the true distribution, and the number of included studies. We compared 15 models implemented in the Frequentist or in the Bayesian framework. We found small differences with respect to bias between the different models but larger differences in the level of coverage probability. In scenarios with large between-study variance, all models were substantially biased in the estimation of the mean treatment effect. This implies that focusing only on the mean treatment effect of random effects meta-analysis can be misleading when substantial heterogeneity is suspected or outliers are present.

# 1. Introduction

Meta-analysis is the statistical combination of the results from two or more individual studies that meet pre-specified eligibility criteria with an aim to answer a specific research question. It generally requires that studies are sufficiently homogeneous to be synthesized. In the presence of heterogeneity, though, a random-effects model may be used[1]. Conventional random-effects meta-analysis assumes that the underlying effects follow a normal distribution, and thus allows for some variability across the available studies[2]. The Cochrane Handbook states that meta-analyses of very diverse studies can be misleading and the presence of heterogeneity affects the extent to which generalizable conclusions can be formed[3,4]. However, it is usually unclear how much heterogeneity may be acceptable and how conclusions may be affected in the presence of significant heterogeneity. This leads to a tendency for meta-analysts to ignore the extent of variation of study results and to focus on the estimated summary effect with its confidence interval only, without realizing that as the variance of the effects' distribution increases the mean becomes less representative of the studies at hand[5,6].

Figure 1 presents random effects meta-analyses of two simulated datasets having binary outcomes in which the study effect measures are log odds ratios. For both datasets, we assumed equal within-study sample sizes and generated the number of events from a binomial distribution (see also Section 3.1). Then, we assigned an underlying normal random effects distribution with a common mean but different variance: $N(0.5, 0.0001)$ and $N(0.5, 2.63)$. The estimated mean in both meta-analyses is 0.52 and, although the confidence interval of the diamond in panel (b) is wider, only marginally crosses the line of no difference. Here, focusing solely on the two diamonds and ignoring the variation between the study-specific effects distributions of these two meta-analyses would probably lead to similar conclusions.

Considerable heterogeneity is not the only situation where assuming a single normal distribution underlying all studies might not capture well enough the structure of the data and conventional meta-analysis practices could be problematic[7]. For instance, the presence of one or more outlying studies is a common phenomenon; in such meta-analyses the use of long-tailed or skewed distributions would possibly describe the structure of the data better[8,9]. In addition, relative effects are frequently associated with one or more study characteristics (i.e. effect modifiers). If these characteristics are available in the data, subgroup analyses or meta-regression are used to investigate whether they differentiate the studies. Nevertheless, there might be effect modifiers which are unknown, unobserved, or unavailable since obtaining data on every variable of interest in a systematic review is usually challenging. In case an important association between study characteristics and their treatment effects is suspected but cannot be investigated, assuming a bimodal or a multimodal distribution for the random effects might be more reasonable than a single normal distribution[10].

Despite the aforementioned limitations of the conventional random effects model, in the vast majority of meta-analyses the between-study variability is modelled through a single normal distribution. Potential reasons for this choice are convenience, model simplicity, tradition, software availability and lack of understanding of the model assumptions. Other more flexible modeling approaches have been suggested in the literature but, to our knowledge, they have rarely been used in clinical applications[11].

In this article, we review and evaluate several meta-analysis models that make different assumptions about the between-study distribution. We first performed a systematic review aiming to identify and summarize all available statistical models for meta-analysis that allow alternative non-normal distributions for the random effects. Subsequently, we conducted a simulation study to compare the identified models and assess their performance under different scenarios. The rest of the article is structured as follows: In Section 2, we briefly describe the methods of our systematic review and we provide an overview of the identified meta-analysis models. Section 3 presents our simulation study and summarizes our findings. In Section 4, we further compare the evaluated models using specific simulated datasets. Finally, in Section 5 we discuss the implications of our findings and in Section 6 we provide concluding remarks.

## 2. Systematic Review

### 2.1 Search and selection of articles

We searched for published articles presenting or evaluating models for meta-analysis that avoid the assumption of a normal distribution for the random effects. The last search was performed on 14 October 2024.

First, we searched in PubMed using the following search algorithm: (meta-analy*[Title] OR synthesi*[Title]) AND (non-normal*[Title/Abstract] OR mixture[Title/Abstract] OR non-parametric*[Title/Abstract] OR flexible random distribution models[Title/Abstract] OR

skewed[Title/Abstract]) AND (model[Title/Abstract] OR approach[Title/Abstract]). Given that some eligible articles might have been published in journals not included in PubMed we further searched in other related journals (such as *Journal of the American Statistical Association, Annals of Statistics,* etc). Finally, we screened the references of the included articles for potentially additional eligible articles.

Eligible articles were those introducing new meta-analysis models, methodological reviews, simulation studies, or commentaries on the properties and characteristics of the models of interest. Overviews of reviews or articles implementing alternative distributions in other parts of the meta-analysis model (e.g. within-study distribution, control group risk, patient-level data distributions) were excluded. We included only articles published in English. Relevant models for meta-analysis of diagnostic test accuracy studies were eligible. Articles about synthesis of gene association studies were excluded.

From each article, we extracted information about the distributional framework(s) proposed or evaluated. We also extracted the meta-analytic setting and the type of data for which the identified models have been suggested. Finally, the theoretical properties and the performance (if available) of the identified models were also extracted.

## 2.2 Search results

We identified 1278 articles through PubMed out of which 1221 were excluded by screening the titles and the abstracts and 36 after reading the full text. Six additional articles that met our inclusion criteria were identified through hand-searching in specific journals. We ended up with 27 eligible articles involving 24 alternative distributions for the random effects[11–37]. The detailed flow chart is available in Supplementary material Figure 1.

## 2.3 Description of the identified models

The identified models can be classified into four main categories based on their between-study distributional assumptions: a) skewed extensions of normal and t-distributions, b) beta distribution c) mixtures of distributions, and d) distributions based on Dirichlet Process priors. In the majority of the articles, the proposed model was constructed under the Bayesian framework. In terms of software, most articles provided code but only a few developed an accompanying R package[12,38–43]. A summary of the characteristics of the eligible articles can be found in Supplementary material Table 1. In the following paragraphs, we start by describing the conventional model where a normal distribution is assumed for the random effects and continue with the description of the alternative models identified through our systematic review.

### 2.3.1 Conventional normal model

Suppose that $Y_1, Y_2, \ldots, Y_n$, are the observed effect sizes for the $n$ studies in the meta-analysis with corresponding underlying effects denoted by $\theta_1, \ldots, \theta_n$. The conventional random effects meta-analysis model assumes for $i = 1, \ldots, n$ that

$$Y_i \sim N(\theta_i, \sigma_i^2)$$
$$\theta_i \sim N(\mu, \tau^2) \quad (1)$$

where $\sigma_i^2$ is the variance of $Y_i$ that usually is assumed known, $\mu$ is the mean of the random effects' distribution (usually referred to as the "summary treatment effect"), and $\tau^2$ is the between-study variance. For brevity, we refer to the model of Equation (1) in the rest of the manuscript simply as the "normal model".

### 2.3.2 t-distribution

The simplest way to allow for some extreme effects in meta-analysis (e.g. outlying studies) is to replace the normal distribution in Equation (1) with a t-distribution[12–14]. In that case,

$$\theta_i \sim t(\mu, \omega, \nu) \quad (2)$$

where $\mu$ and $\tau^2$ are the mean and variance of the t-distribution with scale parameter $\omega^2 = \tau^2 \frac{(\nu-2)}{\nu}$, and $\nu$ is the degrees of freedom determining the weight of the tails. The t-distribution is similar to the normal but it has more weight in the tails and thus outliers generally tend to be less influential. Beath[15] also developed the R package *metaplus*[38] and implemented the above t-distribution model.

A multivariate extension of the t-distribution model has been proposed by Bodnar and Bodnar[16] for meta-analysis of multiple outcomes. Comparing the multivariate normal and the multivariate t-distribution models with several prior distributions in simulations and real data applications, resulted in the multivariate-t model yielding consistently wider credible intervals reflecting the influence of heavy tails. The authors also developed an accompanied R package called *BayesMultMeta*[39].

### 2.3.3 Skewed extensions of normal and t-distribution

To allow for further flexibility and avoid the assumption of a symmetric distribution, we can employ a skew-normal (SN) or a skewed t-distribution (ST)[12,13]. This requires introducing a shape parameter $\gamma$ which regulates the skewness of the distribution[44]. Then, in case of a skew-normal distribution, Equation (1) would be modified into

$$\theta_i \sim \text{SN}(\xi, \omega, \gamma)$$

Considering $\mu$, $\tau^2$, and $a$ as the mean, the variance and the skewness of the skew normal distribution, then the location, scale, and shape parameters, $\xi$, $\omega$, and $\gamma$ respectively, are defined as

$$\xi = \mu - \omega b \delta \quad (1)$$

$$\omega^2 = \frac{\tau^2}{[1-(b\delta)^2]} \quad (2)$$

$$a = \frac{4-\pi}{2} \frac{(b\delta)^3}{(1-(b\delta)^2)^3} \quad (3)$$

where $b = \sqrt{\frac{2}{\pi}}$ and $\delta = \frac{\gamma}{\sqrt{1+\gamma^2}}$. When $a = 0$, the above distribution coincides with the normal distribution. Alternatively, a skewed t-distribution can be used, namely

$$\theta_i \sim \text{ST}(\xi, \omega, \nu, \gamma)$$

where $\xi$, $\omega$, and $\gamma$ are again obtained as a function of the mean, the variance and the skewness of the skewed t-distribution. The above distributions are positively skewed for $\gamma > 0$ and negatively skewed for $\gamma < 0$.

Based on two simulated datasets – one normal and one skewed scenario – and on two real datasets involving some outliers, Lee and Thompson[13] found small differences in the estimation of the mean and the variance of the normal, skew-normal and skew-t distributions. However, relaxing the normality assumption improved model fit and yielded more skewed predictive distributions. They, additionally, provided bivariate extensions of the above models assuming that the treatment effect and the baseline risk are correlated. A bivariate skew-normal model is also suggested by Negeri and Beyene[17] for meta-analysis of diagnostic test accuracy (DTA) studies to model specificity and sensitivity jointly. Both articles conclude that the non-normal models improve model fit and precision when the data are skewed. However, the complexity added by the extra parameters they involve is a key limitation.

### 2.3.4 Other skewed distributions

On top of the skew-normal and skewed t-distribution, Noma et al.[12] proposed the use of three alternative skewed distributions:

- The asymmetric Subbotin distribution (type II)[45,46] being an extension of the symmetric Subbotin distribution, previously proposed for meta-analyses with outliers by Baker and Jackson[14], that can express sharper skewness and excess kurtosis.
- The Jones–Faddy distribution[47] that involves a kurtosis parameter instead of the degrees of freedom.
- The sinh-arcsinh distribution[48] that offers increased flexibility as it can express both symmetric and skewed shapes as well as heavy or light tail-weight.

For a detailed description of these distributions, we refer to the original article[12]. The authors applied the above five skewed distributions to two meta-analyses datasets and compared the results with the normal and the t-distribution models. Given that the skewed distributions provided slightly different mean estimates with narrower credible intervals and resulted in more skewed posterior distributions, they suggest using them as a sensitivity analysis and choosing the most suitable mode based on model fit criteria (such as DIC[49]). They also developed an R package, called *flexmeta*[12], which is linked to the *rstan*[50] package.

### 2.3.5    Beta distribution

Baker and Jackson[14], apart from the t-, Subbotin, and arcsinh distributions, also considered the use of a beta distribution for the random effects restricted on a constraint interval which results to a short-tailed distribution for meta-analyses that completely lack outliers. The proposed between-study distribution is

$$\theta_i \sim \text{Beta}(a_0, b_0)$$

where $a_0 = \mu(\frac{\mu(1-\mu)-\tau^2}{\tau^2})$ and $b_0 = (1-\mu)(\frac{\mu(1-\mu)-\tau^2}{\tau^2})$ with $a_0, b_0 > 1$. The authors compare these models in three meta-analyses with different settings: presence of one outlier, several outliers, and no obvious outliers. They conclude that the use of long-tailed distributions significantly reduces the weight of the outlying studies and might be also more appropriate for meta-analyses where publication bias is suspected.

Chen et al.[18] propose a "hybrid" beta-binomial model for DTA meta-analyses that allows combining case-control and cohort studies. For case-control studies, the random effects follow a bivariate Sarmanov beta distribution[51], accounting for correlations between sensitivity and specificity. For cohort studies, a trivariate Sarmanov beta distribution[51] is used to capture correlations between pairs of sensitivity, specificity, and prevalence. Their models are implemented in the R package *xmeta*[40]. More details can be found in the original article[18].

### 2.3.6 Mixture of distributions

An alternative flexible way to model the random effects is to use a 'mixture' of two or more distributions. This approach might be more relevant when the data seem to naturally come from two or more sub-populations or when several outlying studies are present. Mixture models aim to identify latent subgroups of studies (mixture components) and to estimate each subgroup's mean and variance along with the corresponding mixing proportions. Either different distributions (e.g. a normal distribution and a t-distribution) or the same distribution with different parameters can be employed to the mixture. Hence, a mixture of $k$ normal distributions is

$$\theta_i \sim w_1 \text{N}(\mu_1, \tau_1^2) + \cdots + w_k \text{N}(\mu_k, \tau_k^2) \qquad (6)$$

where $\mu_1, \ldots, \mu_k$ and $\tau_1^2, \ldots, \tau_k^2$ are the subgroup-specific mean and variance for the $k$ subgroups with $w_1, \ldots, w_k$ being the corresponding mixing weights with $\sum_{z=1}^{k} w_z = 1$.

Beath[15] describes a finite mixture model for the random effects for outlier detection. The model considers two normal distributions with common mean ($\mu_c$) and different variances ($\tau_1^2, \tau_2^2$) corresponding to two subgroups of studies representing non-outlying and outlying studies. A bootstrap likelihood ratio test is used to determine whether there are any outliers by comparing models with and without outliers; the outlier studies are identified using posterior predicted probabilities. The weight ($w_1, w_2$) of each distribution in the mixture is proportional to the number of studies in the respective subgroup. Here, Equation (6) becomes

$$\theta_i \sim w_1 \text{N}(\mu_c, \tau_1^2) + w_2 \text{N}(\mu_c, \tau_2^2)$$

and parameters are estimated through an expectation-maximization (EM) algorithm[52]. Then, the summary mean effect is estimated including all studies but with outliers being down-weighted due to the larger variance assumed for their subgroup. An extension of this model incorporating covariates can be used to further explain the observed heterogeneity. Several case studies are provided to point out the importance of identifying and properly modeling outlying studies due to their influence on the estimation of the overall treatment effect. The above model has been implemented in the *metaplus*[38] package in R.

Brown et al.[19] introduce a different two-component normal mixture model where each component is based on a regression model incorporating covariates both at the within- and the between-study level. The use of two components reflects the presence of two suspected subgroups and mixture weights represent the proportion of studies in each subgroup.

Finucane et al.[20] introduce a semi-parametric density estimation meta-analysis model for dependent study effect sizes with covariates. The proposed model is a finite mixture of normal distributions with weights assigned according to the stick and breaking process (see Section 2.3.7). This approach leverages information from both aggregate and individual participant data when available.

Zhang et al.[21] propose a latent mixture-based moderator analysis as a way to disentangle the observed heterogeneity without requiring information on the contributing factors. Specifically, they assume a mixture of several normal distributions for the random effects and then they use an automated data-driven algorithm to decompose the mixture components. They suggest this analysis as a useful step prior to standard moderator analysis (e.g. meta-regression), where researchers can then use the resulting components to examine deeper potential moderating effects.

Eusebi et al.[22] also suggest a similar finite mixture model of bivariate normal distributions and they extend their model to incorporate covariates for predicting latent subgroup classification. They apply their proposed model through the *Latent GOLD 4.5* software[53].

Lopes et al.[23] suggest the use of a mixture of multivariate normal distributions for longitudinal data incorporating a time component as well as other covariates for which they suspect non-linearity. The random effects distribution is decomposed into one part that is common across all studies and a second part that is specific to each study and captures the variability between patients within the same study. This results to a distribution for the random effects which depends on the patients' measurements within studies and on study-level covariates.

Baker and Jackson[24] suggested a similar model to Beath's[15], but expressed the weight of the outlying studies in the mixture as a function of their variance. They also proposed a skewed marginal distribution which is a mixture of a normal and a lagged-normal distribution. The latter is the sum (or difference) of one or more exponential distributions and a normal distribution. This model further includes parameters for skewness and kurtosis. Accounting for covariates is also possible by expressing the proposed model in a regression form. The suggested model appeared to have better fit than the t-distribution model by Baker and Jackson[14], the skewed-t model by Lee and Thompson[13] and the normal one for non-normal data in real data applications.

Sangnawakij et al.[25] suggest a likelihood-based non-parametric mixture model for meta-analyses with rare events that can be used either with arm-based or contrast-based data. They employ a mixture algorithm that assigns study-arms or studies to a fixed number of components with the component parameters being estimated via Poisson regression. Mixture weights are defined as the proportion of studies in each component. The algorithm generates estimates from all possible data classifications to the mixture components. This model was first proposed by Böhning et al.[26] along with a bivariate extension, but without considering arm-based data.

Van Houwelingen et al[27] introduced the use of another (EM) algorithm[54] that results in a discrete mixture distribution for the random effects. They also proposed a bivariate extension of this approach assuming random effects for both treatment and control arms in order to investigate their relationship with the overall mean treatment effect.

Karabatsos et al.[28] propose a Bayesian infinite random-intercept mixture of regressions. This is, in practice, a discrete mixture model where the random intercept parameter is derived from a covariate-dependent infinite mixture distribution. This model allows for a wide range of distributions for the random effects, including unimodal symmetric, skewed, or multimodal distributions. The proposed model can identify which of the included covariates may be important predictors. Based on a meta-analysis of highly heterogeneous studies involving 24 covariates and multiple study reports, the authors suggested that the proposed non-parametric model describes better the distribution of the underlying treatment effects in comparison to various versions of normal fixed and random effects models. It was also considered to fit better to the data based on goodness of fit measures. They applied their proposed model through a software[55] developed by Karabatsos[55].

A flexible finite mixture model of bivariate normal distributions is proposed by Schlattmann et al.[29] for DTA meta-analysis. The model uses a bivariate version of the model of Equation (6) to model sensitivity and specificity simultaneously. This model was applied using the *CAMAN*[56] and *mada*[57] R packages developed by Schlattmanna et al.[29] and Doebler[57] respectively.

### 2.3.7 Dirichlet Process Priors

A further possibility to model the random effects distribution is through a class of nonparametric priors, namely Dirichlet Process (DP) priors. The use of DP priors is the most flexible option among the identified alternative models and offers the opportunity to automatically identify the potential underlying clustering of the data: here relevant subgroups of studies. In practice, the DP is a generalization of the Dirichlet distribution with the property that, for any finite partition of the parameter space, the DP marginalizes to a Dirichlet distribution[58,59].

Muthukumarana and Tiwari[30] suggested a simple DP model for meta-analysis described as

$$\theta_i \sim F$$
$$F \sim \text{DP}(\alpha, F_0)$$
$$F_0 \sim N(\mu_b, \tau_b^2) \qquad (7)$$

where $F_0$ is the base distribution that controls the mean of the process and can be any distribution; here a normal distribution with mean $\mu_b$ and variance $\tau_b^2$ is assumed. The concentration parameter $\alpha \geq 0$ measures the variability of $F$ around $F_0$, with higher values of $\alpha$ suggesting that $F$ is 'closer' to $F_0$. It can be given either a fixed value or a prior distribution and larger values (e.g. larger than the number of studies) give more weight to the base distribution.

A DP can be implemented using different approaches, such as the Chinese restaurant process[60,61], the Polya urn scheme[62], or the stick and breaking process[63]. For example, the latter generates a set of $x_j \sim F_0$ points (i.e. location parameters), and their corresponding weights $p_j$ which depend on the value of the concentration parameter $\alpha$. Then, $F = \sum_{j=1}^{\infty} p_j I_{x_j}(x)$, where $I_{x_j}$ is an indicator variable with $I_{x_j}(x_j) = 1$ and $I_{x_j}(x) = 0$ otherwise. The weights $p_j$ are defined using a recursive scheme that repeatedly samples from the beta distribution $\text{Beta}(1, a)$. That is, $p_j = q_j * \prod_{i=1}^{j-1}(1 - q_i)$, where $q_j \sim \text{Beta}(1, a)$. A truncation that allows obtaining a plausible approximation to F is usually applied to make the process faster; for instance, the number of studies in the meta-analysis was used here as a truncation point.

Based on a real data example, Muthukumarana and Tiwari[30] suggest that their proposed method provides narrower credible intervals for the study specific effects in comparison to the normal model. In their simulation study, under highly heterogeneous and non-normal scenarios, the DP model had a better fit to the data compared with the normal model.

Ohlssen et al.[31] also suggest the use of a truncated DP (TDP) by truncating at a maximum number of mass points $N$. Hence, the prior for $F$ in Equation (7) now becomes

$$F \sim \text{TDP}(\alpha, F_o, N)$$

Here, in contrast to Muthukumarana and Tiwari[30], $N$ is closely related to the concentration parameter $\alpha$ through $N \approx 1 - \alpha \log_e \varepsilon$ with $\varepsilon$ representing the expected value of the probability assigned to the final point, $E[p_N]$. In practice, using a very small value for $\varepsilon$, such as 0.01, would give

$$N \approx 1 + 5\alpha \qquad (8)$$

Several articles, though, use $N = n$[30,34]. The above model assumes a discrete random effects distribution by implementing a mixture of points (DPMp) but it can be extended to a continuous random effects distribution by implementing a mixture of (normal) distributions instead (DPMd). Using a meta-analysis of routinely collected data as well as a simulation study with data generated from normal and mixture of binomial distributions, the authors suggest that the truncated DP models fit the data better than the normal one and correctly identify clusters among the underlying effects. Their simulations also imply that the estimated value for the concentration parameter $\alpha$ is an indicator of whether the random effects distribution is normal or not.

A modified version of the DPMd model suggested by Ohlssen et al.[31] is presented by Burr and Doss[11,32]. Specifically, they introduce the "conditional DP" by replacing Equation (7) with $F \sim \text{DPMd}^{\mu_b}(\alpha, F_0)$; that is the conditional distribution for $F$ given that the posterior median of $F$ is $\mu_b$. This model might be preferable when the number of studies is small. It also has the advantage that the estimation of $\mu_b$ is not influenced materially by the presence of few outlying studies. Using these two models for an exemplar meta-analysis allowed to identify subgroup differences in one single analysis. The authors also developed an R package, called *bspmma*[41], in which the conditional and non-conditional DPM models have been implemented. Although the package runs the models fast, it only supports the normal distribution as base distribution ($F_0$) and lacks flexibility in prior distributions.

Jo et al.[33] describe a flexible mixed effects meta-regression model to properly handle aggregate data for several subpopulations. They also employ a DPMd model for the random effects which can be alternatively written as an infinite mixture of truncated normal distributions.

Cao et al.[34] recently extended the truncated DPMp model described by Ohlssen et al.[31] for the estimation of reference intervals for a new individual or a new study[64]. The authors built their models using *Nimble*[65], an R package which, although it is not specific to meta-analysis, contains automated functions for the implementation of different DP processes such as the stick and breaking process or the Chinese restaurant process. More details can be found in the original article[34].

Another extension of the DP priors is proposed by Dunson et al.[35] who introduce the "Matrix Stick-Breaking Process" (MSBP) designed for individual participant data meta-analyses with several predictors per study. In practice, this model modifies the stick and breaking process to allow borrowing information across predictors and studies simultaneously. This is achieved by the incorporated increased probability of two studies being clustered together for a specific predictor given that those studies have already been clustered together for other predictors. Simulated examples indicated some superiority of MSBP compared to other DP models in terms of MSE only in cases with moderate or large number of coefficients.

Branscum and Hanson[36] introduce a Polya-tree mixture model for meta-analysis. Polya-tree priors can be seen as a generalization to DP priors[66–69]; they can be discrete or continuous with the latter resulting in less distinct cluster effects. The Polya-tree's model structure resembles the "conditional DP" model described by Burr and Doss[11,32] but the process weights depend on the tree partition of the space. In a simulation study comparing the proposed model and the normal model with data generated from a skewed bimodal distribution, the former resulted in posterior distributions closer to the distribution of the true effects. The Polya-tree mixture model has been implemented in the *DPpackage*[42] in R through the *PTmeta* function. However, the package only allows a base distribution with $\mu_b = 0$. The same package allows fitting the DPMp and DPMd models through the functions *DPmeta* and *DPMmeta* respectively.

Finally, Barrientos et al.[37] recently proposed a TDP for network meta-analysis aiming to identify treatment hierarchies or equalities. Two different base distributions are considered: a normal and a 'spike and slab'; the latter is a two-component mixture distribution. The spike component is a spread distribution typically centered at 0 while the slab component could be either a disintegrated distribution at 0 or a continuous distribution centered at 0 with small variance. This model has been implemented in the R package *CBnetworkMA*[43].

## 3. Simulation Study

We compared the normal model with some of the alternative models identified through our systematic review in a simulation study which we present following the recommendations by Morris et al[70].

### 3.1 Data generating mechanism

We generated a range of meta-analysis datasets consisting of $n$ studies comparing an active with a control intervention for a dichotomous outcome. Based on empirical data[71], we set $n = 14, 26$ to represent meta-analyses of moderate and large size respectively. We generated the underlying effects of the studies $\theta_i$ $(i = 1, ..., n)$ from a normal distribution, a skew-normal, or a mixture of two normal distributions: $N(\mu, \tau^2)$, $SN(\xi, \omega, \gamma)$ and $w_1 N(\mu_1, \tau_1^2) + w_2 N(\mu_2, \tau_2^2)$ with $w_1 = 0.3$ and $w_2 = 0.7$ respectively. We considered scenarios with $\mu = 0, 0.5$ and $\mu_1 = 0, \mu_2 = 1$ to represent the absence and the presence of a treatment effect. We also considered $\tau^2 = 0.12, 2.63$ to reflect scenarios with moderate and high between-study variance respectively based on the empirical distributions for log odds ratios provided by Turner et al[72] for subjective outcomes and comparisons between a pharmacological intervention and a placebo/control. For the mixture of two normal distributions, we used $\tau_1^2 = 0.12$ and $\tau_2^2 = 0.005, 0.12, 2.63$. The parameters $\xi, \omega, \gamma$ were derived from Equations (3), (4) and (5) assuming $a = 0.79$.

Then, we generated arm-level data from a discrete uniform distribution ranging from 50 to 500 assuming equal within-study sample sizes for the treatment ($m_{it}$) and the control group ($m_{ic}$). To generate the number of events in the control group, we used $c_i \sim \text{Bin}(m_{ic}, \rho_{ic})$ with $\rho_{ic} \sim U(0.05, 0.65)$ and for the events of the treatment group $t_i \sim \text{Bin}(m_{it}, \rho_{it})$ with $\rho_{it} = \frac{\rho_{ic} e^{\theta_i}}{1 - \rho_{ic} + \rho_{ic} e^{\theta_i}}$[73]. Studies generated with zero events in both arms were excluded from the respective meta-analysis; two studies only from two different scenarios had to be excluded.

Overall, we assessed 22 scenarios and for each scenario we generated 1000 meta-analyses. A summary of all scenarios is available in Table 1.

### 3.2 Evaluated models and software used

We evaluated 15 different models; 11 implemented in the Bayesian framework and 4 in the Frequentist framework (Table 2).

For each Bayesian model we used a Binomial likelihood for the arm-level data and we applied one of the following four different between-study distributions, each with different prior specifications for their parameters:
1. normal with a non-informative normal prior $N(0, 10^4)$ for $\mu$ and for $\tau$
    a. a half-normal $HN(0,1)$, or

b.  a uniform prior U(0,10)
2. t-distribution with a N(0,10⁴) prior for $\mu$ and an exponential prior Exp(0.10)[12] for the degrees of freedom ($\nu$) combined, as above, with (a) or (b) priors for the scale parameter ($\omega$)
3. skew-normal with a N(0, 10⁴) prior for $\xi$ and a normal prior N(0,25) for the shape parameter $\gamma$ combined again with (a) or (b) priors for the scale parameter ($\omega$)
4. a truncated DPMp (constructed using the stick and breaking process) with a N(0, 10⁴) prior for $\mu_b$ combined with (a) or (b) priors for $\tau_b$ as well as for the concentration parameter ($\alpha$)
   i.   a uniform prior U(0.3,5) resulting from Equation (8) in a truncation parameter $N = 26$, using the prior's maximum value $\alpha = 5$, or
   ii.  a U(0.3,10) prior resulting in $N = 51$ for $\alpha = 10$. This allows for a larger number of potential clusters to be created and a better approximation to the full process[74], or
   iii. Gamma prior $\Gamma(1,1)$ on $\alpha$ and $N = n$[30,34]

Under the Frequentist framework, we used one model with binomial likelihood on the arm-level data and normal between-study distribution with a maximum likelihood (ML) estimator for variance and three models with normal approximation to the observed log-odds ratios of the studies with different between-study distributions:

1. normal with the restricted maximum likelihood (REML) estimator for $\tau$
2. t-distribution (normal-t model) with ML estimator for $\tau$ and profile-likelihood confidence intervals[75]
3. mixture of two normal distributions with common mean and different variances (common-mean mixture) as described by Beath[15] again with ML estimator for $\tau$

We ran all simulations in the R statistical software version 4.2.2 (October 31, 2022)[76]. Bayesian models were built in JAGS[77] and Stan[78] through the *rjags*[79], *R2jags*[80] and *cmdstanr*[81] packages. For all analyses, we ran 2 chains with 50,000 iterations and a 10,000 burn-in period. The convergence of all Bayesian models was checked by extracting the ratio of between-chain to within-chain variance $\hat{R}$ with $\hat{R} < 1.05$ suggesting convergence. Datasets that did not reach convergence were excluded from the simulation results but the respective model and scenario was included, given that at least 95% of the datasets converged. The Frequentist normal-normal and binomial-normal models were run using the *metafor* package[82], while the t-distribution and common-mean mixture models were run with the *metaplus* package[38].

### 3.3 Estimands and performance measures

The estimands of interest are the mean and variance of the random effect distribution and the shrinkage estimates of the studies measured on the log scale. The latter were only available for the eleven models fitted under the Bayesian framework and for the normal-normal(REML) model. For the common-mean mixture model, that provides two different variance estimates (for outlying and for non-outlying studies), we used in our main results the variance for outlying studies as a more conservative option. We also extracted and monitored the skewness parameter from the skew-normal model and three additional parameters from the DP model: the concentration parameter as well as the mean and the variance of the base distribution $F_0$. The clustering assignment given by the DP was investigated in three meta-analyses from the simulation.

The performance of the models was investigated in terms of mean bias defined as the absolute mean difference between the estimated and the true parameters of interest averaged over the simulated datasets in each scenario. We further calculated the mean coverage defined as the percentage of the corresponding 95% confidence or credible intervals that included the true values of the parameters. The confidence interval for the coverage was calculated as $0.95 \pm 1.96\sqrt{\frac{0.95(1-0.95)}{1000}}$. Additionally, the different models were compared in terms of the mean squared error (MSE). Finally, for $\tau^2$ we also calculated the percentage bias defined as the mean bias divided by the true $\tau^2$ and the normalized MSE defined as the MSE divided by the square of the true $\tau^2$.

### 3.4 Simulation Results
#### 3.4.1 Mean treatment effect

Figure 2 shows the results in terms of average bias for the mean of the random effects distribution across all scenarios and models investigated. Supplementary material Table 2 shows the 44 model-scenario pairs that we excluded due to lack of convergence of that model for a specific scenario (i.e., a model with at least 50 failures out of 1000 simulated datasets for a specific scenario). Overall, we found small differences in the average bias between the different models for the estimated mean treatment effect. Irrespective of the true distribution of the random effects, the bias of all models was relatively small for scenarios with moderate between-study variance but substantially higher for scenarios with large between-study variance. The bias tended to be slightly smaller for scenarios with more studies. The normal-t and the common-mean mixture models tended to perform worse than the other models for scenarios with large between-study variance. When data were generated from a skew-normal distribution with large variance and in some mixture scenarios the binomial-SN(HN) model seems to have the best and most stable performance along with the normal-normal(REML) and binomial-normal(ML) models. Under all scenarios considered, the DP models that reached convergence performed similarly to the normal models. This is possibly due to the use of a normal base distribution. Similar trends were observed in terms of the MSE, where the normal-normal(REML), the binomial-normal(ML) and the binomial-normal models with HN prior for $\tau$ or $\tau_b$ generally yielded lower MSE in scenarios with large between-study variance (Supplementary material Figure 2).

In terms of coverage probability, most of the models performed poorly in scenarios with large between-study variance except the binomial-normal and binomial-DP models using a Uniform prior for $\tau$ and $\tau_b$ respectively that were close to the nominal level (Figure 3). As expected, in four out of the six mixture scenarios where the shape of the true distribution was clearly bimodal (Scenarios 17, 18, 20, 21), the DP models that reached convergence had the best performance.

### 3.4.2 Between-study variance

In all investigated scenarios with moderate between-study variance the bias of all models regarding the estimation of the overall $\tau^2$ was generally small with the common-mean mixture model being usually the most biased (Figure 4a). This is likely due to the choice of using the variance from outlying studies, which inflates the estimate when only a few outliers are present, and does not properly reflect the random effects variance (Supplementary material Table 3). The bias was substantially higher in scenarios with large between-study variance where the binomial models with a HN prior for $\tau$ or $\tau_b$ appeared to be the least biased especially for scenarios with 14 studies. Again, the binomial-SN(HN) model had the best performance in skew normal scenarios. Similar patterns are observed regarding the MSE of the different models (Supplementary material Figure 3a).

However, when considering the percentage bias of $\tau^2$, we observe small differences between models across all scenarios and in some cases the percentage bias for moderate between-study variance is larger than that for large between-study variance (Figure 4b). Similar findings are obtained when using the normalized MSE of $\tau^2$ (Supplementary material Figure 3b). In terms of coverage probability (Supplementary material Figure 4), we could only compare the eleven models fitted in the Bayesian framework and the normal-normal(REML) model. In all normal and skew-normal scenarios, the normal models and the binomial-DP models using a Uniform prior for $\tau_b$ outperformed the others.

### 3.4.3 Underlying effects of the studies

Figure 5 presents the mean bias of the study-specific effects averaged within and across meta-analyses for each scenario. In the two mixture scenarios with two clearly distinct distributions (scenarios 17 and 20), the DP models which reached convergence, estimated the study-specific effects with the smallest bias on average but they are equally or more biased than the other models in all other scenarios. Differences between the other models are small. Again, the average bias of all models is higher for scenarios with large between-study variance.

Additional results from our simulation study are available in the Supplementary material Table 4 and Supplementary material Table 5.

## 4. Selected simulated datasets

To investigate the performance of the different models on additional aspects, we extracted three datasets from our simulation generated from a normal (scenario 6), a skew-normal (scenario 13), and a mixture (scenario 20) distribution, that were the most representative of each case. At each of these datasets we applied the four Bayesian models with the best performance in terms of coverage probability and bias for the mean treatment effect $\mu$. Hence, we selected:

- the binomial-normal(HN), the binomial-t(HN), the binomial-SN(HN) and the binomial-DP-n(Unif/Gamma) for scenario 6.
- the binomial-normal(HN), the binomial-t(Unif), the binomial-SN(Unif) and binomial-DP-26(HN/Unif) for scenario 13.
- the binomial-normal(HN), the binomial-t(HN), the binomial-SN(HN) and the binomial-DP-51(HN/Unif) for scenario 20.

Overall, all models gave similar estimates for the study-specific estimates and the mean treatment effect estimate for all three datasets (Supplementary material Figure 5, Supplementary material Figure 6 and Supplementary material Figure 7). The DP models tended to give more precise estimates for the effects of some studies, which in some cases resulted in 95% credible intervals not including the study-specific true values. They also tended to produce estimates of study-specific effects closer to the mean of the cluster they belong which is expected as highlighted by Burr and Doss[11,32].

For the dataset generated from the normal distribution, all models yielded similarly biased estimates of $\mu$. The $\tau^2$ was underestimated by all models with the binomial-t(HN) model providing the closest estimation to the true value but with the widest credible interval. The binomial-SN(HN) model produced the narrowest prediction interval. The DP model here struggled to identify potential clusters and suggested a minimum of 4 to 8 clusters. This is in line with the estimate of 3.99 for the concentration parameter, which is large enough to support a normal random effects distribution given that the $\Gamma(1,1)$ prior used is rather restricted around zero and one. For all models, despite the high probability of a positive mean treatment effect, the probability of a new study with $\theta_i < 0$ is 0.45. This aligns with the clusters' probabilities generated by the DP model, indicating that a new study has nearly equal chances of being assigned to either the first or the second cluster, with corresponding means of -0.70 and 0.28, respectively (Supplementary material Table 6). Supplementary material Figure 9 shows that all models have a good overlap between the distribution of the true effects and the posterior distribution of the random effects.

For the dataset generated from the skew-normal distribution, the point estimates from all models were similar for the mean treatment effect but more biased than for the dataset assuming normal random effects and with larger probabilities for $\mu < 0$ (Supplementary material Figure 6). Specifically, the credible interval of the binomial-t(Unif) model only marginally included the true mean value of zero. As expected, the binomial-SN(Unif) model is the least biased in the estimation of $\mu$, also producing the narrowest prediction interval. Additionally, the estimated mean skewness parameter from this model is 0.85 which is close to the true skewness of 0.79. The DP model suggests the presence of three clusters with only study 22 occupying the third cluster; this implies that this study is a potential outlier contributing to the skewness of the data (Supplementary material Figure 8(b)). Also, the concentration parameter was estimated as 2.78 supporting the presence of a non-normal random effects' distribution. The $Pr(\mu_{new} < 0)$ ranges from 61% to 66% using the parametric models indicating that most likely a new study would fall below the null effect. This is also supported by the DP model, where the first cluster has the larger number of studies and a corresponding negative mean of -0.30 (Supplementary material Figure 6). The DP model produces a highly peaked posterior distribution, which leads to a seemingly poor overlap with the distribution of the true effects (Supplementary material Figure 10). However, this is possibly due to the use of a discrete distribution for the random effects rather than a continuous one.

Similar estimates between the models were also obtained for the dataset generated from the mixture distribution regarding the mean treatment effect and the between-study variance with the skew normal and the DP model estimating accurately the mean treatment effect and only the binomial-t(HN) model resulting in a slightly higher estimate (Supplementary material Figure 7). All the models gave zero probability of $\mu < 0$, while the prediction interval from the binomial-SN(HN) model was the narrowest. The DP model identified three clusters with the first and the second cluster being very close. The first cluster is mainly formed due to study 6 which, though, can also be assigned to the second cluster with the same probability. Study 16 is assigned to the third cluster, but it can also be assigned to the second one with the same probability. Clusters 2 and 3 have means -0.06 and 0.99, namely very close to the true means of the two distributions in the mixture. The concentration parameter was estimated 3.90 which is small enough (given the prior $U(0,10)$) to support the presence of a non-normal distribution. The probability of a new study falling into the third cluster is the highest compared to the other two clusters, indicating that the effect of a new study is likely to be above 0 and closer to the third cluster's mean of 0.99 (Supplementary material Table 6). This conclusion is also supported by the parametric models since the $Pr(\mu_{new} < 0)$ is around 0.10. Finally, as expected, the posterior distribution of the random effects from the DP model has the best overlap with the distribution of the true study effects for this dataset (Supplementary material Figure 11).

## 5. Discussion

In the present article, we identified and compared several meta-analysis models that relax the between-study normality assumption which is typically used in published meta-analyses. We first performed a methodological systematic review to search for alternative random effects models suggested in the literature and then conducted a simulation study to investigate their performance under different scenarios. We found 27 eligible articles suggesting 24 different random effects distributions: based on long-tail or skewed extensions of the normal and t-distribution, on the beta distribution, on mixtures of two or more distributions, and on variations of DP priors. In our simulation, we generated meta-analyses with binary data and we considered 22 scenarios varying the true distribution, the level of between-study variance, the true mean treatment effect, and the number of studies. We compared 11 of the identified alternative models between them as well as with 4 versions of the normal model.

Our findings highlight the limited ability of all evaluated models to accurately estimate the mean treatment effect as the true between-study variance increases. This has important implications for applications of meta-analysis since, to date, the focus in the literature is most often on the mean treatment effect and its confidence interval even when the results of the studies differ substantially. In such cases, exploring the factors that may cause between-study variance is of great importance. Mixture and semi-parametric models can give insight on the underlying clustering of the studies and assist to form homogenous subgroups that may share common characteristics.

Overall, our simulation results suggest that using a HN prior for variance parameters tended to improve the performance of the Bayesian models in terms of bias and MSE across all scenarios, while the use of Uniform prior resulted in coverage probabilities closer to the nominal level. We found minor differences among the evaluated models in the bias of the estimated mean treatment effect which was materially increased for scenarios with large between-study variance. These findings imply that in meta-analyses with small to moderate between-study variance the estimated mean treatment effect from any model, including the normal models, may provide a sufficiently accurate summary of the studies at hand, whereas in meta-analyses with large between-study variance focusing only on the estimated mean treatment effect may lead to meaningless and/or spurious conclusions.

Similar results were obtained for the absolute bias of the between-study variance. However, the respective percentage bias was not affected by the value of the true variance. We did not find important differences between the different Bayesian models in the average bias of the estimated study-specific effects in scenarios with $\tau^2 = 0.12$. In scenarios with large between-study variance, though, and when the true distribution was closer to normal or skew normal the DP models performed worse than the other models, whereas when the true distribution was clearly bimodal the DP models performed substantially better. Hence, the use of such complex models seems mostly beneficial when the presence of a bimodal or multimodal distribution is suspected. This is often the case when effect modifiers operate across studies.

In the three selected simulated datasets, we obtained similar patterns as in the full simulation study for the bias of the mean treatment effect. For the skew-normal dataset, the skew normal model was the least biased for the estimation of between-study variance, while the DP model correctly identified three clusters. For the mixture dataset, both the skew normal and the DP models performed similarly well with the latter accurately identifying two to three clusters. In all three datasets, the use of additional parameters and statistics, such as the concertation parameter, the prediction intervals, and the probability for the presence of an effect, helped us to understand better the structure of the data and the variation of the study effects.

Of course, our study is not free of limitations. First, we did not compare all the identified alternative models in our simulation; nevertheless, to our knowledge, this is the most extensive simulation study evaluating several meta-analysis models assuming different between-study distributions on a wide range of scenarios. Our results are in agreement with previous smaller simulation studies[13,14,30,31], but most of these studies did not consider the level of between-study variance as a characteristic affecting the performance of the models. Using scenarios with more studies might have improved the performance of some of the models and particularly the DP models where the underlying distribution is treated as a random variable. We selected the number of studies, though, based on empirical data of meta-analyses from different medical fields involving observational studies to generate meta-analyses often encountered in the literature. Future simulation studies may also consider meta-analyses with very few or a lot of studies encountered in certain situations (e.g. meta-analyses including RCTs only, meta-analyses in social science). We should acknowledge, that some of the models considered here might not be applicable or useful in the presence of very few studies. Including studies with rare events could be also something interesting to explore in future work.

In addition, we only used a fixed skewness parameter ($a = 0.79$) which might have not resulted in many highly skewed datasets. Varying the skewness parameter in the data generating process could also potentially provide more insights into the performance of the skew normal models. Moreover, the performance of the t-models cannot be fully assessed in the present study due to the fact that we did not generate data from a t-distribution. Our results imply, though, that the t-models used here are not flexible enough to be used for datasets with different underlying distributions. The DP models with a Uniform prior on $\tau_b$ failed to reach convergence for most scenarios with low to moderate between-study variance. This might be due to the relatively small number of studies that we used in the data generation. Using non-normal distributions for the base distribution of the DP models and the DPMd approach would be interesting directions for future research. Finally, for the common-mean mixture model, using the variance of outlying studies in the simulation results might have influenced its performance since this model appeared the most biased under all scenarios.

# 6. Conclusions

Our systematic review revealed that several alternative flexible meta-analyses models have been suggested in the literature. Despite the potential advantages these models may have, they are typically not used in clinical applications. This might be due to lack of user-friendly software or lack of understanding of the model assumptions. Overall, we found small differences in the performance measures of our simulation between the models we compared. The normal-normal(REML) model, which is the most commonly used model in clinical applications, did not appear to be inferior than the other more sophisticated and complex models for most scenarios. Our results imply that when substantial heterogeneity among studies is suspected or outlying studies are present, making inferences solely based on the mean treatment effect of random effects meta-analysis could be misleading since it might be substantially biased and it does not consider the variation of the study effects. In such cases, identifying the factors that differentiate the studies and looking at the prediction intervals are more informative than the estimated mean of the random-effects distribution. We encourage meta-analysts to always examine carefully the plausibility of the between-study normality assumption and the extent of between-study variability before undertaking their analysis. Sensitivity analysis using different models may give some insight into the underlying structure of the data and help understanding factors causing between-study variance.

# Figures

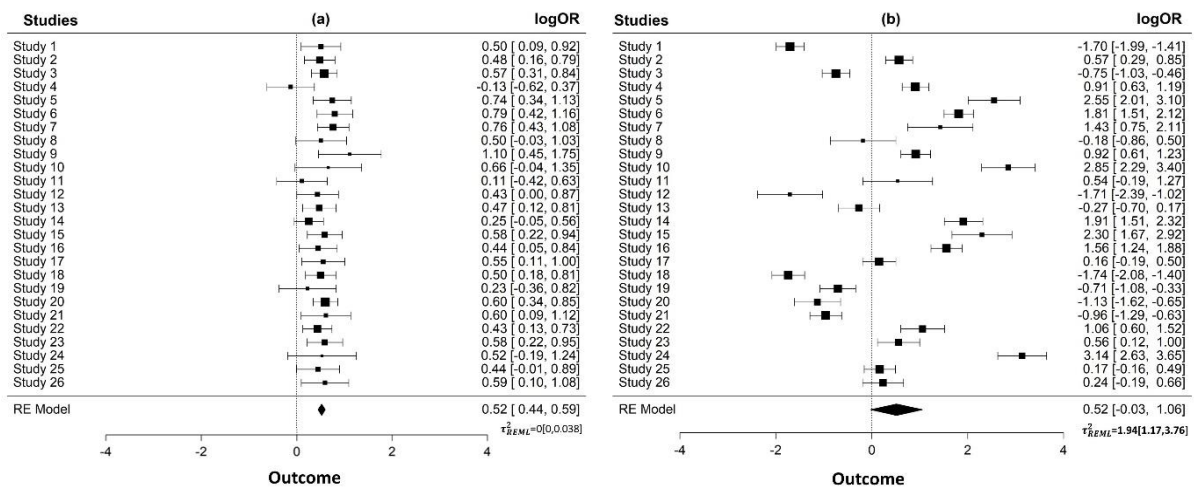

**Figure 1**. Meta-analysis of two simulated sets of studies generated from two normal distributions with the same mean but different variances. The dataset of panel (a) was generated from $N(0.5, 0.0001)$ and that of panel (b) from $N(0.5, 2.63)$.

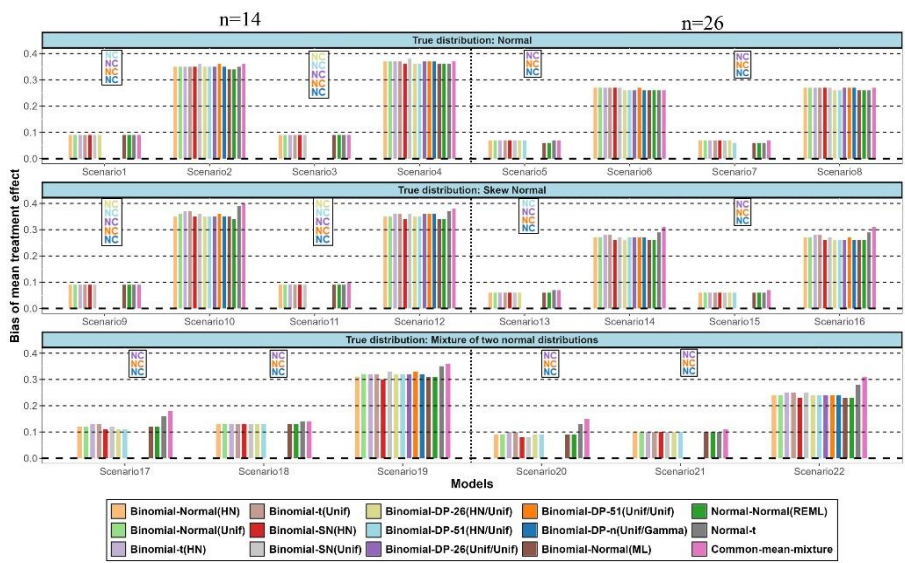

**Figure 2**. Simulation results in terms of mean bias for the mean of the random effects distribution. The names of the models are explained in Table 2. (NC=Non-convergence)

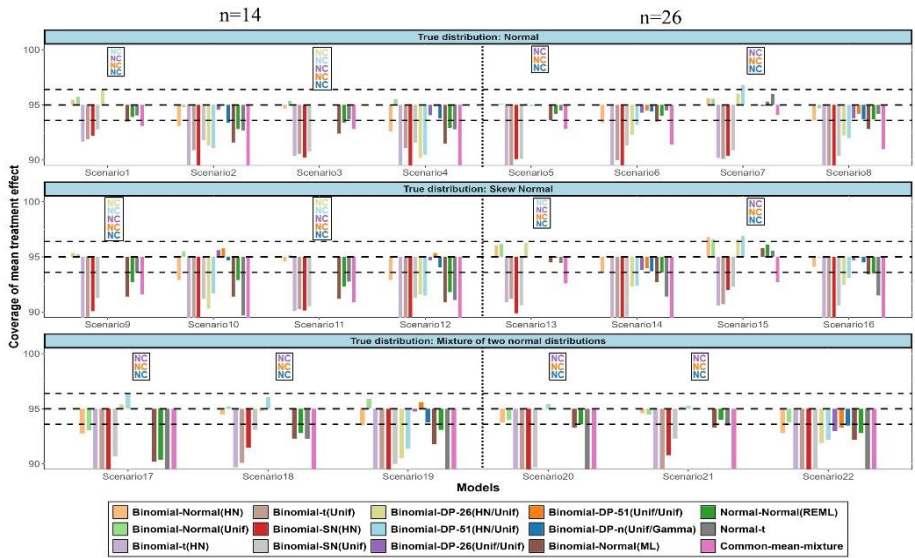

**Figure 3**. Simulation results in terms of coverage probability for the mean of the random effects distribution. The horizontal lines represent the upper and lower bounds of the 95% confidence interval for the nominal level. The names of the models are explained in Table 2. (NC=Non-convergence)

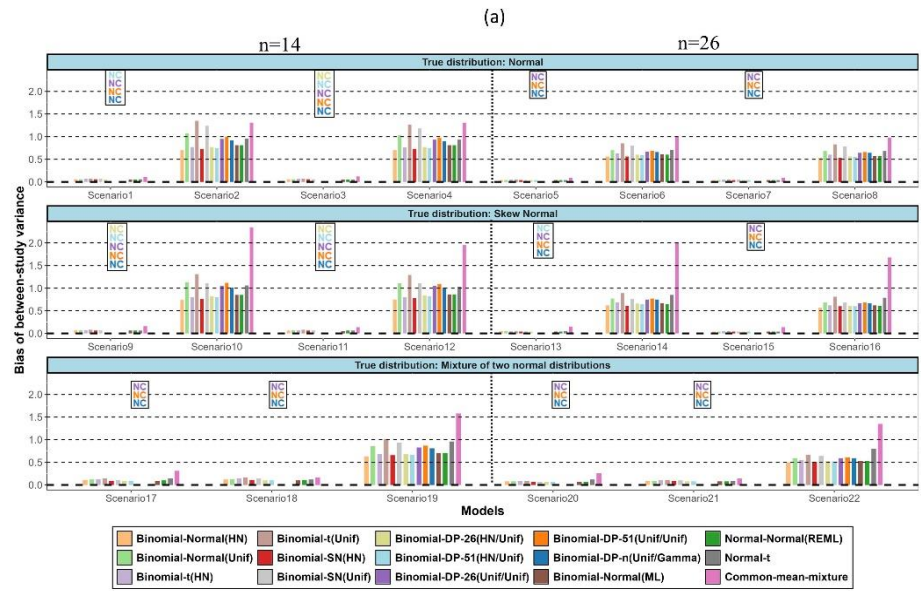

(a)

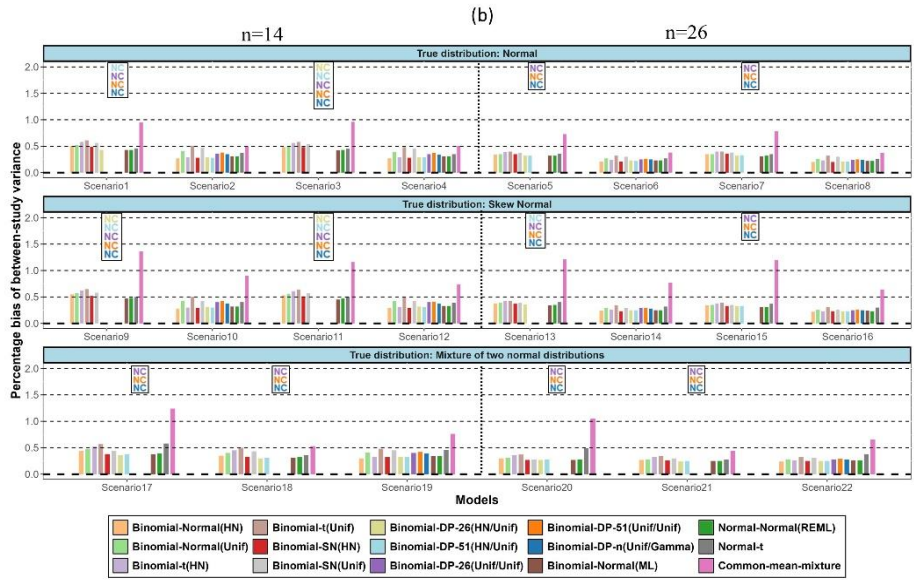

**Figure 4**. Simulation results in terms of mean of absolute bias (a) and percentage bias (b) for the between-study variance. The names of the models are explained in Table 2. (NC=Non-convergence)

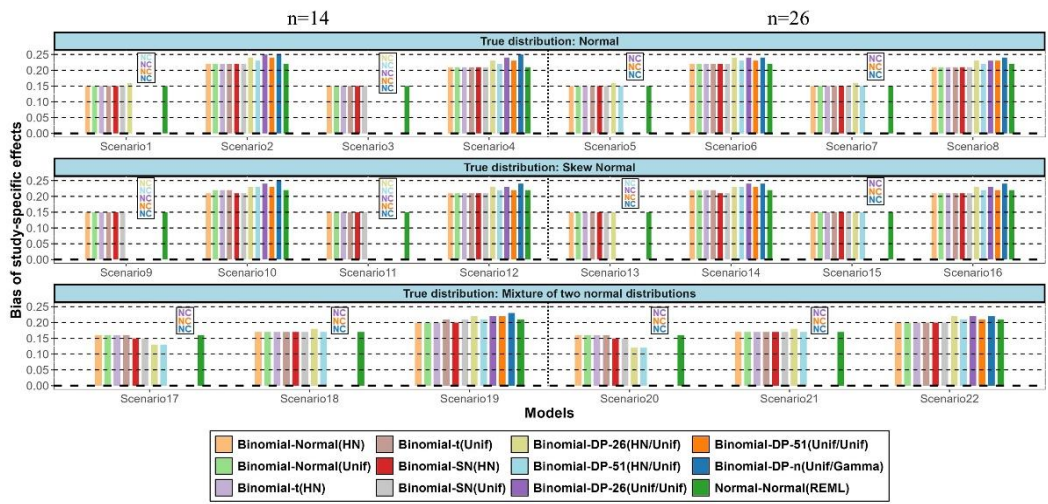

**Figure 5**. Simulation results in terms of mean bias for the study-specific treatment effects averaged within meta-analyses and across meta-analyses. The names of the models are explained in Table 2. (NC=Non-convergence)

**Supplementary material Figures**

**Supplementary material Figure 1**. Flow chart of the article selection process.

**Supplementary material Figure 2**. Simulation results in terms of mean square error for the mean of the random effects distribution. The names of the models are explained in Table 2. (NC=Non-convergence)

**Supplementary material Figure 3**. Simulation results in terms of mean square error (a) and normalized mean square error (b) for the between-study variance. The names of the models are explained in Table 2. (NC=Non-convergence)

**Supplementary material Figure 4**. Simulation results in terms of coverage probability for the between-study variance. The horizontal lines represent the upper and lower bounds of the 95% confidence interval for the nominal level. The names of the models are explained in Table 2. (NC=Non-convergence)

**Supplementary material Figure 5**. Estimated study-specific effects of the selected simulated dataset from Scenario 6 using the binomial-Normal(HN) (panel (a)), the binomial-t(HN) (panel (b)), the binomial-SN(HN) (panel (c)), and the binomial-DP-n(Unif/Gamma) (panel (d)) models. The diamonds represent the estimated mean treatment effect from each model. The probabilities of the mean treatment effect or a new study ($\mu_{new}$) to be less than 0 are also presented. In panel (d), the two extra columns give the cluster assignment and the probability for each study of belonging to the respective cluster. The dashed vertical lines represent the means of the identified clusters. Red studies are those belonging to the dominant cluster and the black ones those belonging to the other clusters.

**Supplementary material Figure 6**. Estimated study-specific effects of the selected simulated dataset from Scenario 13 using the binomial-Normal(HN) (panel (a)), the binomial-t(Unif) (panel (b)), the binomial-SN(Unif) (panel (c)), and the binomial-DP-26(HN/Unif) (panel (d)) models. The diamonds represent the estimated mean treatment effect from each model. The probabilities of the mean treatment effect or a new study ($\mu_{new}$) to be less than 0 are also presented. In panel (d), the two extra columns give the cluster assignment and the probability for each study of belonging to the respective cluster. The dashed vertical lines represent the means of the identified clusters. Black, red and blue studies are those belonging to the first, second and third cluster respectively.

**Supplementary material Figure 7**. Estimated study-specific effects of the selected simulated dataset from Scenario 20 using the binomial-Normal(HN) (panel (a)), the binomial-t(HN) (panel (b)), the binomial-SN(HN) (panel (c)), and the binomial-DP-51(HN/Unif) (panel (d)) models. The diamonds represent the estimated mean treatment effect from each model. The probabilities of the mean treatment effect or a new study ($\mu_{new}$) to be less than 0 are also presented. In panel (d), the two extra columns give the cluster assignment and the probability for each study of belonging to the respective cluster. The dashed vertical lines represent the means of the identified clusters. Blue, red, and black studies are those belonging to the first, second, and third cluster respectively.

**Supplementary material Figure 8**. Distribution of the true study-specific effects of the selected simulated dataset from Scenario 6 (panel (a)), Scenario 13 (panel (b)), and Scenario 20 (panel (c)).

**Supplementary material Figure 9**. Overlap between the distribution of the true-study-specific effects (Scenario 6) and the posterior distribution of the random effects from each Bayesian model used.

**Supplementary material Figure 10.** Overlap between the distribution of the true-study-specific effects (Scenario 13) and the posterior distribution of the random effects from each Bayesian model used.

**Supplementary material Figure 11**. Overlap between the distribution of the true-study-specific effects (Scenario 20) and the posterior distribution of the random effects from each Bayesian model used.

**References**

1. Borenstein M, Hedges LV, Higgins JPT, et al. A basic introduction to fixed-effect and random-effects models for meta-analysis. *Res Synth Methods* 2010; 1: 97–111.

2. Riley RD, Higgins JPT, Deeks JJ. Interpretation of random effects meta-analyses. *BMJ* 2011; 342: d549.

3. Higgins JPT, Thomas J, Chandler J, et al. *Cochrane Handbook for Systematic Reviews of Interventions version 6.4*. 2nd ed. Chichester (UK): John Wiley & Sons, 2019.

4. Deeks JJ, Higgins JPT, Altman DG. Chapter 10: Analysing data and undertaking meta-analyses. In: Higgins JPT, Thomas J, Chandler J, et al. (eds) *Cochrane Handbook for Systematic Reviews of Interventions version 6.4*. 2nd ed. Chichester (UK): John Wiley & Sons, 2019.

5. Imrey PB. Limitations of Meta-analyses of Studies With High Heterogeneity. *JAMA Netw Open* 2020; 3: e1919325.

6. Schroll JB, Moustgaard R, Gøtzsche PC. Dealing with substantial heterogeneity in Cochrane reviews. Cross-sectional study. *BMC Med Res Methodol* 2011; 11: 22.

7. Wang CC, Lee WC. Evaluation of the Normality Assumption in Meta-Analyses. *Am J Wang Epidemiol* 2020; 189: 235–242.

8. Higgins JPT, White IR, Anzures-Cabrera J. Meta-analysis of skewed data: combining results reported on log-transformed or raw scales. *Stat Med* 2008; 27: 6072–6092.

9. Meng Z, Wang J, Lin L, et al. Sensitivity analysis with iterative outlier detection for systematic reviews and meta-analyses. *Stat Med* 2024; 43: 1549–1563.

10. Jackson D, White IR. When should meta-analysis avoid making hidden normality assumptions? *Biom J* 2018; 60: 1040–1058.

11. Burr D, Doss H, Cooke GE, et al. A meta-analysis of studies on the association of the platelet PlA polymorphism of glycoprotein IIIa and risk of coronary heart disease. *Stat Med* 2003; 22: 1741–1760.

12. Noma H, Nagashima K, Kato S, et al. Meta-analysis Using Flexible Random-effects Distribution Models. *J Epidemiol* 2022; 32: 441–448.

13. Lee KJ, Thompson SG. Flexible parametric models for random-effects distributions. *Stat Med* 2008; 27: 418–434.

14. Baker R, Jackson D. A new approach to outliers in meta-analysis. *Health Care Manag Sci* 2008; 11: 121–131.

15. Beath KJ. A finite mixture method for outlier detection and robustness in meta-analysis. *Res Synth Methods* 2014; 5: 285–293.

16. Bodnar O, Bodnar T. Objective Bayesian Meta-Analysis Based on Generalized Marginal Multivariate Random Effects Model. *Bayesian Anal* 2024; 19: 531–564.

17. Negeri ZF, Beyene J. Skew-normal random-effects model for meta-analysis of diagnostic test accuracy (DTA) studies. *Biom J* 2020; 62: 1223–1244.

18. Chen Y, Liu Y, Chu H et al. A simple and robust method for multivariate meta-analysis of diagnostic test accuracy. *Stat Med* 2017; 36: 105–121.

19. Brown CH, Wang W, Sandler I. Examining How Context Changes Intervention Impact: The Use of Effect Sizes in Multilevel Mixture Meta-Analysis. *Child Dev Perspect* 2008; 2: 198–305.

20. Finucane MM, Paciorek CJ, Stevens GA, et al. Semiparametric Bayesian Density Estimation With Disparate Data Sources: A Meta-Analysis of Global Childhood Undernutrition. *J Am Stat Assoc* 2015; 110: 889–901.

21. Zhang N, Wang M, Xu H. Disentangling effect size heterogeneity in meta-analysis: A latent mixture approach. *Psychol Methods* 2022; 27: 373–399.

22. Eusebi P, Reitsma JB, Vermunt JK. Latent class bivariate model for the meta-analysis of diagnostic test accuracy studies. *BMC Med Res Methodol* 2014; 14: 88.

23. Lopes HF, Müller P, Rosner GL. Bayesian meta-analysis for longitudinal data models using multivariate mixture priors. *Biometrics* 2003; 59: 66–75.

24. Baker R, Jackson D. New models for describing outliers in meta-analysis. *Res Synth Methods* 2016; 7: 314–328.

25. Sangnawakij P, Böhning D, Holling H et al. Nonparametric estimation of the random effects distribution for the risk or rate ratio in rare events meta-analysis with the arm-based and contrast-based approaches. *Stat Med* 2024; 43: 706–730.

26. Böhning D, Martin S, Sangnawakij P et al. Nonparametric Estimation of Effect Heterogeneity in Rare Events Meta-Analysis: Bivariate, Discrete Mixture Model. *Lobachevskii Journal of Mathematics* 2021; 42: 308–317.

27. Van Houwelingen HC, Zwinderman KH, Stijnen T. A bivariate approach to meta-analysis. *Stat Med* 1993; 12: 2273–2284.

28. Karabatsos G, Talbott E, Walker SG. A Bayesian nonparametric meta-analysis model. *Res Synth Methods* 2015; 6: 28–44.

29. Schlattmann P, Verba M, Dewey M, et al. Mixture models in diagnostic meta-analyses-clustering summary receiver operating characteristic curves accounted for heterogeneity and correlation. *J Clin Epidemiol* 2015; 68: 61–72.

30. Muthukumarana S, Tiwari RC. Meta-analysis using Dirichlet process. *Stat Methods Med Res* 2016; 25: 352–365.

31. Ohlssen DI, Sharples LD, Spiegelhalter DJ. Flexible random-effects models using Bayesian semi-parametric models: applications to institutional comparisons. *Stat Med* 2007; 26: 2088–2112.

32. Burr D, Doss H. A Bayesian Semiparametric Model for Random-Effects Meta-Analysis. *J Am Stat Assoc* 2005; 100: 242–251.

33. Jo S, Park B, Chung Y, et al. Bayesian semiparametric mixed effects models for meta-analysis of the literature data: An application to cadmium toxicity studies. *Stat Med* 2021; 40: 3762–3778.

34. Cao W, Chu H, Hanson T, et al. A Bayesian nonparametric meta-analysis model for estimating the reference interval. *Stat Med* 2024; 43: 1905–1919.

35. Dunson DB, Xue Y, Carin L. The Matrix Stick-Breaking Process. *J Am Stat Assoc* 2008; 103: 317–327.

36. Branscum AJ, Hanson TE. Bayesian nonparametric meta-analysis using Polya tree mixture models. *Biometrics* 2008; 64: 825–833.

37. Barrientos AF, Page GL, Lin L. Non-parametric Bayesian approach to multiple treatment comparisons in network meta-analysis with application to comparisons of anti-depressants. J R Stat Soc C Appl Stat. 2024; 73: 13331–354.

38. Beath KJ. metaplus: An R Package for the Analysis of Robust Meta-Analysis and Meta-Regression. *R J* 2016; 8: 5.

39. Bodnar O, Bodnar T, Thorsén E. BayesMultMeta: Bayesian Multivariate Meta-Analysis, 2022 URL https://CRAN.R-project.org/ package=BayesMultMeta, R package version 0.1.1.

40. Hong C, Luo C, Tong G et al. xmeta: A Toolbox for Multivariate Meta-Analysis, 2023. URL https://CRAN.R-project.org/package=xmeta, R package version 1.3.2.

41. Burr D. bspmma: An R Package for Bayesian Semiparametric Models for Meta-Analysis. *J Stat Softw* 2012; 50: 1–23.

42. Jara A, Hanson TE, Quintana FA, et al. DPpackage: Bayesian Non- and Semi-parametric Modelling in R. *J Stat Softw* 2011; 40: 1–30.

43. Page G. CBnetworkMA: Contrast-Based Bayesian Network Meta Analysis, 2024. URL https://CRAN.R-project.org/package=CBnetworkMA, R package version 0.1.0.

44. Azzalini A. Frontmatter. In: *The Skew-Normal and Related Families*. Cambridge: Cambridge University Press, 2013.

45. Azzalini A, Capitanio A. Distributions Generated by Perturbation of Symmetry with Emphasis on a Multivariate Skew t-Distribution. *J R Stat Soc B*.2003; 65: 367–389.

46. Azzalini A. A Class of Distributions Which Includes the Normal Ones. *Scand J Stat* 1985; 12: 171–178.

47. Jones MC, Faddy MJ. A Skew Extension of the T-Distribution, with Applications. *J R Stat Soc B.* 2003; 65: 159–174.

48. Jones MC, Pewsey A. Sinh-arcsinh distributions. *Biometrika* 2009; 96: 761–780.

49. Spiegelhalter DJ, Best NG, Carlin BP et al. Bayesian measures of model complexity and fit (with discussion). *J R Stat Soc B.* 2002; 64: 583–639.

50. Stan Development team. Rstan: the R interface to stan, 2024, URL http://mc-stan.org/, R package version 2.32.6.

51. Ting MLT. Properties and applications of the sarmanov family of bivariate distributions. *Commun Stat - Theory Methods* 1996; 25: 1207–1222.

52. McLachlan G, Peel D. *Finite Mixture Models*. Wiley, New York: Wiley Series in Probability and Statistics, 2000.

53. Vermunt JK, Magidson J. *LG-Syntax user's guide: Manual for Latent GOLD 4.5 Syntax module*. Belmont: Statistical Innovations Inc, 2008.

54. Dempster AP, Laird NM, Rubin DB. Maximum Likelihood from Incomplete Data Via the *EM* Algorithm. *J R Stat Soc B.* 1977; 39: 1–22.

55. Karabatsos G. A menu-driven software package of Bayesian nonparametric (and parametric) mixed models for regression analysis and density estimation. *Behav Res Methods*. 2017; 49: 335–362.

56. Schlattmann P, Hoehne J. CAMAN: Finite Mixture Models and Meta-Analysis Tools - Based on C.A.MAN, 2023, URL http://CRAN.R-project.org/package=CAMAN, R package version 0.78.

57. Doebler P. Meta-Analysis of Diagnostic Accuracy with mada, 2022, URL http://CRAN.R-project.org/package=mada, R package version 0.5.11.

58. Sammut C, Webb GI. *Encyclopedia of machine learning and data mining*. 2nd ed. NY:Springer New York, 2017.

59. Li Y, Schofield E, Gönen M. A tutorial on Dirichlet Process mixture modeling. *J Math Psychol* 2019; 91: 128–144.

60. Kim JH, Kim D, Kim S, et al. Modeling topic hierarchies with the recursive chinese restaurant process. In: *Proceedings of the 21st ACM international conference on Information and knowledge management* 2012; 783–792.

61. Blei DP, Griffiths TL, Jordan MI. The nested chinese restaurant process and bayesian nonparametric inference of topic hierarchies. *J ACM* 2010; 57: 1–30.

62. Blackwell D, MacQueen JB. Ferguson distributions via Polya urn schemes. *Ann Statist* 1973; 1: 353–355.

63. Ishwaran H, James LF. Gibbs Sampling Methods for Stick-Breaking Priors. *J Am Stat Assoc* 2001; 96: 161–173.

64. Wright EM, Royston P. Calculating reference intervals for laboratory measurements. *Stat Methods Med Res* 1999; 8: 93–112.

65. De Valpine P, Turek D, Paciorek CJ, et al. Programming With Models: Writing Statistical Algorithms for General Model Structures With NIMBLE. *J Comput Graph Stat* 2017; 26: 403–413.

66. Lavine M. Some Aspects of Polya Tree Distributions for Statistical Modelling. *Ann Statist* 1992; 20: 1222–1235.

67. Lavine M. More aspects of Polya tree distributions for statistical modelling. *Ann Statist* 1994; 22: 1161–1176.

68. Hanson TE. Inference for mixtures of finite Polya tree models. *J Am Stat Assoc* 2006; 101: 1548–1565.

69. Mauldin RD, Sudderth WD, Williams SC. Polya Trees and Random Distributions. *Ann Statist* 1992; 20: 1203–1221.

70. Morris TP, White IR, Crowther MJ. Using simulation studies to evaluate statistical methods *Stat Med* 2019; 38: 2074–2102.

71. Cheurfa C, Tsokani S, Kontouli KM, et al. Synthesis methods used to combine observational studies and randomised trials in published meta-analyses. *Syst Rev* 2024; 13: 70.

72. Turner RM, Davey J, Clarke MJ, et al. Predicting the extent of heterogeneity in meta-analysis, using empirical data from the Cochrane Database of Systematic Reviews. *Int J Epidemiol* 2012; 41: 818–827.

73. Petropoulou M, Mavridis D. A comparison of 20 heterogeneity variance estimators in statistical synthesis of results from studies: a simulation study. *Stat Med* 2017; 36: 4266–4280.

74. Ishwaran H, Zarepour M. Markov chain Monte Carlo in approximate Dirichlet and beta two-parameter process hierarchical models. *Biometrika* 2000; 87: 371–390.

75. Y. Pawitan. *In All Likelihood: Statistical Modelling and Inference using Likelihood.* Oxford: Clarendon Press, 2001.

76. R Core Team *R: A language and environment for statistical computing* SEP. Vienna, Austria: R Foundation for Statistical Computing, 2022. https://www.R-project.org/.

77. Plummer M. JAGS: A Program for Analysis of Bayesian Graphical Models using Gibbs Sampling. Vienna:*3rd International Workshop on Distributed Statistical Computing, 2003, 1-10.*

78. Carpenter B, Gelman A, Hoffman MD, et al. Stan: A Probabilistic Programming Language. *J Stat Softw* 2017; 76: 1–32.

79. Plummer M. rjags: Bayesian Graphical Models using MCMC, 2023. URL https://CRAN.R-project.org/package=rjags, R package version 4–15.

80. Su YS, Yajima M. R2jags: Using R to Run 'JAGS', 2021. URL https:// CRAN.R-project.org /package=R2jags, R package version 0.7–1.1.

81. Gabry J, Češnovar R. cmdstanr: R Interface to 'CmdStan', 2024. URL https://mc-stan.org/cmdstanr/.

82. Viechtbauer W. Conducting Meta-Analyses in R with the metafor Package. *J Stat Softw* 2010; 36: 1–48

**Table 1.** Summary of the 22 scenarios considered in the simulation study.

| Scenario | True distribution | | | | | | | | | Number of studies (n) |
|---|---|---|---|---|---|---|---|---|---|---|
| | Type | Mean ($\mu$) | Variance ($\tau^2$) | Skewness ($a$) | Location ($\xi$) | Scale ($\omega$) | Shape ($\gamma$) | Weights ($w_z$) | Shape | |
| 1 | Normal | 0 | 0.12 | | | | | | 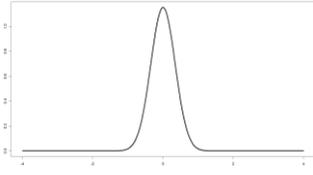 | 14 |
| 2 | Normal | 0 | 2.63 | | | | | | 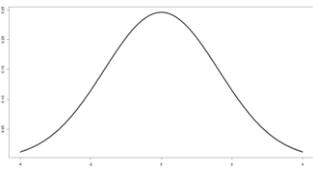 | 14 |
| 3 | Normal | 0.5 | 0.12 | | | | | | 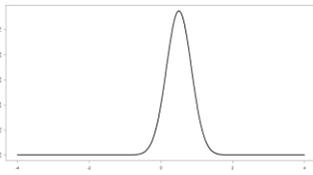 | 14 |
| 4 | Normal | 0.5 | 2.63 | | | | | | 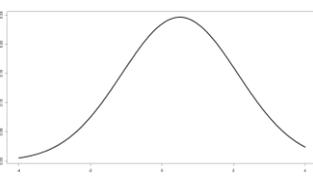 | 14 |
| 5 | Normal | 0 | 0.12 | | | | | | 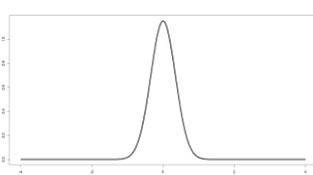 | 26 |

| | | | | | | | | | |
|---|---|---|---|---|---|---|---|---|---|
| 6 | Normal | 0 | 2.63 | | | | | 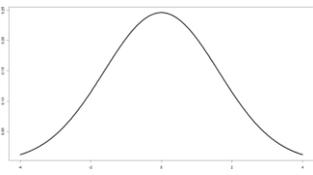 | 26 |
| 7 | Normal | 0.5 | 0.12 | | | | | 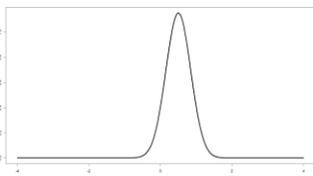 | 26 |
| 8 | Normal | 0.5 | 2.63 | | | | | 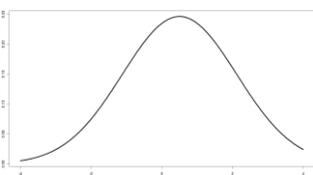 | 26 |
| 9 | Skew-normal | 0 | 0.12 | 0.785 | -0.42 | 0.55 | 4 | 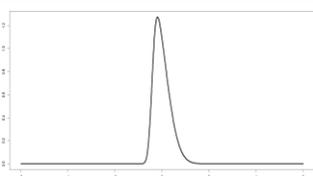 | 14 |
| 10 | Skew-normal | 0 | 2.63 | 0.785 | -1.98 | 2.56 | 4 | 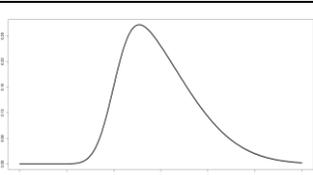 | 14 |
| 11 | Skew-normal | 0.5 | 0.12 | 0.785 | 0.08 | 0.55 | 4 | 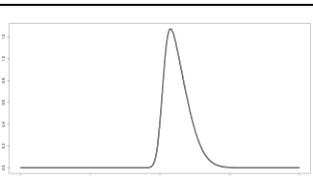 | 14 |

| | | | | | | | | | | |
|---|---|---|---|---|---|---|---|---|---|---|
| 12 | Skew-normal | 0.5 | 2.63 | 0.785 | -1.48 | 2.56 | 4 | | 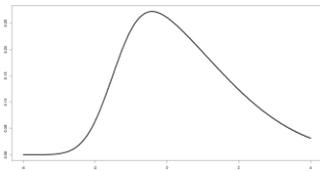 | 14 |
| 13 | Skew-normal | 0 | 0.12 | 0.785 | -0.42 | 0.55 | 4 | | 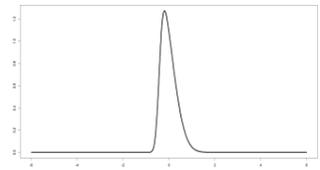 | 26 |
| 14 | Skew-normal | 0 | 2.63 | 0.785 | -1.98 | 2.56 | 4 | | 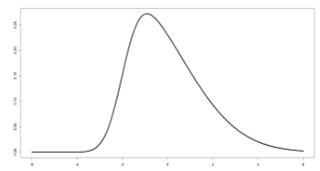 | 26 |
| 15 | Skew-normal | 0.5 | 0.12 | 0.785 | 0.08 | 0.55 | 4 | | 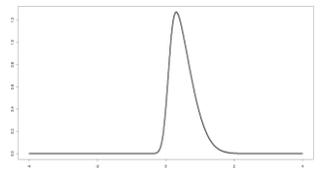 | 26 |
| 16 | Skew-normal | 0.5 | 2.63 | 0.785 | -1.48 | 2.56 | 4 | | 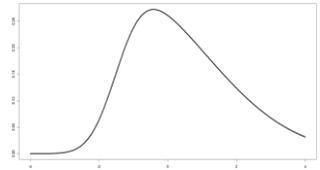 | 26 |
| 17 | Mixture of 2 normal distributions | $\mu_1 = 0$ $\mu_2 = 1$ | $\tau_1^2 = 0.12$ $\tau_2^2 = 0.005$ | | | | | $w_1 = 0.3$ $w_2 = 0.7$ | 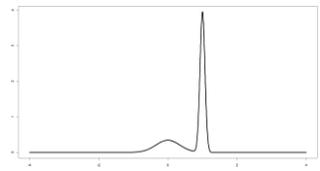 | 14 |

| 18 | Mixture of 2 normal distributions | $\mu_1 = 0$ $\mu_2 = 1$ | $\tau_1^2 = 0.12$ $\tau_2^2 = 0.12$ | | | | | $w_1 = 0.3$ $w_2 = 0.7$ | 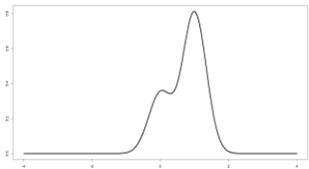 | 14 |
|---|---|---|---|---|---|---|---|---|---|---|
| 19 | Mixture of 2 normal distributions | $\mu_1 = 0$ $\mu_2 = 1$ | $\tau_1^2 = 0.12$ $\tau_2^2 = 2.63$ | | | | | $w_1 = 0.3$ $w_2 = 0.7$ | 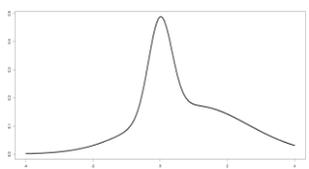 | 14 |
| 20 | Mixture of 2 normal distributions | $\mu_1 = 0$ $\mu_2 = 1$ | $\tau_1^2 = 0.12$ $\tau_2^2 = 0.005$ | | | | | $w_1 = 0.3$ $w_2 = 0.7$ | 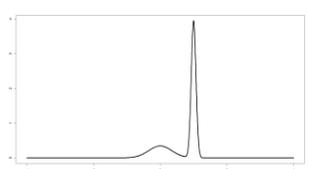 | 26 |
| 21 | Mixture of 2 normal distributions | $\mu_1 = 0$ $\mu_2 = 1$ | $\tau_1^2 = 0.12$ $\tau_2^2 = 0.12$ | | | | | $w_1 = 0.3$ $w_2 = 0.7$ | 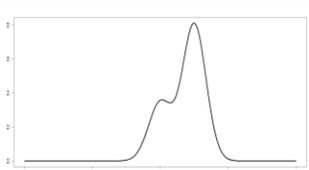 | 26 |
| 22 | Mixture of 2 normal distributions | $\mu_1 = 0$ $\mu_2 = 1$ | $\tau_1^2 = 0.12$ $\tau_2^2 = 2.63$ | | | | | $w_1 = 0.3$ $w_2 = 0.7$ | 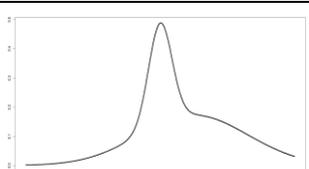 | 26 |

**Table 2.** Summary of the evaluated models in the simulation study.

| Model abbreviation | Framework fitted | Within-study distribution | Between-study distribution | Prior distributions for key parameters | | | | | | | | | DP truncation points (*N*) |
|---|---|---|---|---|---|---|---|---|---|---|---|---|---|
| | | | | Mean ($\mu$) | Between-study standard deviation ($\tau$) | base mean ($\mu_b$) | base variance ($\tau_b$) | concentration ($\alpha$) | location ($\xi$) | Scale ($\omega$) | Shape ($\gamma$) | degrees of freedom ($\nu$) | |
| Binomial-Normal (HN) | Bayesian | Binomial | Normal | N(0, $10^4$) | HN(0,1) | - | - | - | - | - | - | - | - |
| Binomial-Normal (Unif) | Bayesian | Binomial | Normal | N(0, $10^4$) | U(0,10) | - | - | - | - | - | - | - | - |
| Binomial-t(HN) | Bayesian | Binomial | t-distribution | N(0, $10^4$) | - | - | - | - | - | HN(0,1) | - | Exp(0.10) | - |
| Binomial-t(Unif) | Bayesian | Binomial | t-distribution | N(0, $10^4$) | - | - | - | - | - | U(0,10) | - | Exp(0.10) | - |
| Binomial-SN(HN) | Bayesian | Binomial | Skew Normal | - | - | - | - | - | N(0, $10^4$) | HN(0,1) | N(0,25) | - | - |
| Binomial-SN(Unif) | Bayesian | Binomial | Skew Normal | - | - | - | - | - | N(0, $10^4$) | U(0,10) | N(0,25) | - | - |
| Binomial-DP-26 (HN/Unif) | Bayesian | Binomial | DPMp | - | - | N(0, $10^4$) | HN(0,1) | U(0.3,5) | - | - | - | - | 26 |
| Binomial-DP-51 (HN/Unif) | Bayesian | Binomial | DPMp | - | - | N(0, $10^4$) | HN(0,1) | U(0.3,10) | | | | | 51 |
| Binomial-DP-26 (Unif/Unif) | Bayesian | Binomial | DPMp | - | - | N(0, $10^4$) | U(0,10) | U(0.3,5) | - | - | - | - | 26 |
| Binomial-DP-51(Unif/Unif) | Bayesian | Binomial | DPMp | - | - | N(0, $10^4$) | U(0,10) | U(0.3,10) | - | - | - | - | 51 |
| Binomial-DP-n (Unif/Gamma) | Bayesian | Binomial | DPMp | - | - | N(0, $10^4$) | U(0,10) | $\Gamma$(1,1) | - | - | - | - | *n* |
| Binomial-normal(ML) | Frequentist | Binomial | Normal | - | - | - | - | - | - | - | - | - | - |
| Normal-Normal(REML) | Frequentist | Normal | Normal | - | - | - | - | - | - | - | - | - | - |
| Normal-t | Frequentist | Normal | t-distribution | - | - | - | - | - | - | - | - | - | - |
| Normal-Common-mean mixture | Frequentist | Normal | Mixture of 2 normal distributions | - | - | - | - | - | - | - | - | - | - |

| | | | with common mean | | | | | | | | | | |
|---|---|---|---|---|---|---|---|---|---|---|---|---|---|

**Supplementary material Tables**

**Supplementary material Table 1.** Characteristic of all the eligible articles identified through the systematic review.

**Supplementary material Table 2.** Bayesian models and scenarios included (✓) or excluded (✘) from the simulation study due to lack of convergence in more than 5% of the datasets.

**Supplementary material Table 3.** Summary of performance metrics for the common-mean mixture model, reporting results separately for outlying and non-outlying studies. The metrics include mean bias, percentage bias, mean squared error (MSE), and normalized MSE, for both variance components of the model.

**Supplementary material Table 4.** Mean bias of the estimated skewness parameter ($a$) from the skew-normal model in scenarios where data were generated from a skew-normal distribution.

**Supplementary material Table 5.** Estimated means with credible intervals of the concentration parameter ($\alpha$) from the five DP models for the different simulation scenarios.

**Supplementary material Table 6.** Cluster probabilities for each selected simulated dataset using the respective DP model.